\documentclass[amssymb,prb,twocolumn,floats,amsmath,showpacs,showkeys]{revtex4-1}

\usepackage{bm}
\usepackage{graphicx}
\usepackage{epstopdf}

\begin{document}


\title{Ultra low density of CdTe quantum dots grown by MBE }
%

\author{J.~Kobak}
\affiliation{Institute of Experimental Physics, Faculty of Physics, University of Warsaw,
ul. Ho\.za 69, 02-681 Warsaw Poland}

\author{J-G.~Rousset}
\affiliation{Institute of Experimental Physics, Faculty of Physics, University of Warsaw,
ul. Ho\.za 69, 02-681 Warsaw Poland}

\author{R.~Rudniewski}
\affiliation{Institute of Experimental Physics, Faculty of Physics, University of Warsaw,
ul. Ho\.za 69, 02-681 Warsaw Poland}

\author{E.~Janik}
\affiliation{Institute of Experimental Physics, Faculty of Physics, University of Warsaw,
ul. Ho\.za 69, 02-681 Warsaw Poland}

\author{T.~S{\l}upi\'{n}ski}
\affiliation{Institute of Experimental Physics, Faculty of Physics, University of Warsaw,
ul. Ho\.za 69, 02-681 Warsaw Poland}

\author{P.~Kossacki}
\affiliation{Institute of Experimental Physics, Faculty of Physics, University of Warsaw,
ul. Ho\.za 69, 02-681 Warsaw Poland}

\author{A.~Golnik}
\affiliation{Institute of Experimental Physics, Faculty of Physics, University of Warsaw,
ul. Ho\.za 69, 02-681 Warsaw Poland}

\author{W.~Pacuski}
\email{Wojciech.Pacuski@fuw.edu.pl}
\affiliation{Institute of Experimental Physics, Faculty of Physics, University of Warsaw,
ul. Ho\.za 69, 02-681 Warsaw Poland}

%
\begin{abstract}
This work presents methods of controlling the density of self-assembled CdTe quantum dots (QDs) grown by molecular beam epitaxy. Two approaches are discussed: increasing the deposition temperature of CdTe and the reduction of CdTe layer thickness. Photoluminescence (PL) measurements at low temperature confirms that both methods can be used for significant reduction of QDs density from 10$^{10}$QD/cm$^2$ to 10$^7$-10$^8$QD/cm$^2$. For very low QDs density, identification of all QDs lines observed in the spectrum is possible.
\end{abstract}



\maketitle

\section{Introduction}

 The density of QDs in a sample is the crucial limiting factor for individual QDs spectroscopy \cite{Brown2004, Dousse2010, Rodel2012}. Typically II-VI systems, such as self-assembled CdTe/ZnTe QDs, exhibit such a high density of dots\cite{Karczewski1999,Tinjod2003, HSLee2009} (order of 10$^{10}$QD/cm$^2$) that the typical laser spot in microscope system (${\phi}=2$~${\mu}$m) excites hundreds of a dots simultaneously. Good spectral separation of single QD is possible only for dots with particularly low emission energy \cite{Suffczynski2006, Gaj2009, HSLee2009, Kobak2011, Pimpinella2011}. In order to reduce the number of observed QDs one can use masks \cite{Warburton2000,Zrenner2000}, etch mesa structures \cite{Bayer2000}, or use near field spectroscopy \cite{Brun2002,Pimpinella2011} but all such techniques affect the optical properties such as polarization. A much better solution is to find growth conditions which result in the formation of very low density QDs.
This is the motivation of our study. We received good hints from Wojnar et al.\cite{Wojnar2007}, who showed that QDs density can be reduced by thermal annealing (but still hundreds of QDs per laser spot were observed). We developed growth conditions which result in ultra low density of QDs e.g. 1 -  3 QDs per laser spot.
Our preliminary results have shown that CdTe QDs density depends on growth temperature, therefore we systematically investigated the influence of deposition temperature of CdTe layer and ZnTe cap layer on the density of QDs. In order to distinguish between the effect of thermal annealing and effect of CdTe desorption, we also made a reference series of samples grown with various thicknesses of the CdTe layer from which dots were formed.


\section{Growth}

\begin{figure}[!h]
\includegraphics[width=\linewidth]{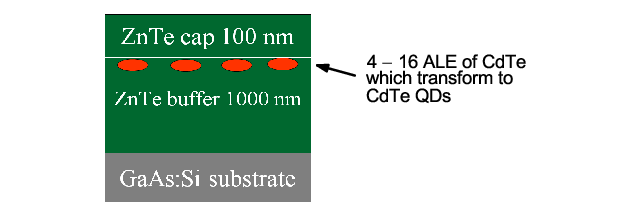}
\centering \caption[]{(color online) Structure of samples - CdTe/ZnTe QDs on GaAs (100) substrates.} \label{fig:sampleStructure}
\end{figure}

Samples were grown using molecular beam epitaxy (MBE) in growth chamber model SVT-35. We used GaAs:Si (100) substrates covered by 1 ${\mu}$m thick ZnTe buffer layers. The scheme of the structure is shown in Fig.~\ref{fig:sampleStructure}.

\begin{figure}[!h]
\includegraphics[width=\linewidth]{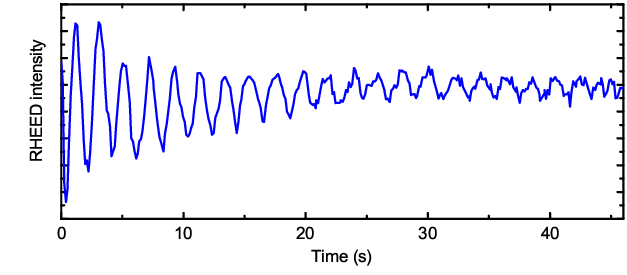}
\centering \caption[]{(color online) RHEED signal intensity oscillations observed after growth interruption of ZnTe buffer.} \label{fig:RHEEDoscillations}
\end{figure}

In order to verify that our ZnTe buffer is smooth the intensity of RHEED (Reflection High Energy Electron Diffraction) signal was recorded as a function of time. After growth interruption, well-defined RHEED signal oscillations (Fig.~\ref{fig:RHEEDoscillations}) due to ZnTe growth were present for a relatively long time (about 50 s) what was considered as a proof of good-quality ZnTe buffer. The buffer was grown always at the same substrate temperature T = 365${^\circ}$C.
Since in our system the thermocouple is in a radiative thermal contact with the substrate, the substrate temperature calibration was obtained by determination of sublimation rate of CdTe and comparing it with data from Ref. \cite{Tatarenko1994}, where thermocouple was in contact with the substrate. This allowed us to determine the relation between the temperature shown by our thermocouple and real substrate temperature. Consequently, for the temperature range 300-400${^\circ}$C the precision of our substrate temperature measurement is ${\pm}$ 5${^\circ}$C.

\begin{figure}[!h]
\includegraphics[width=\linewidth]{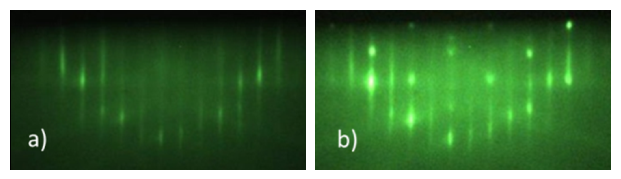}
\centering \caption[]{(color online) Typical RHEED image of CdTe layer a) before and b) after desorption of amorphous tellurium which lead to characteristic transformation from 2D to 3D image.} \label{fig:RHEEDimage}
\end{figure}

CdTe QDs were formed using a method of amorphous tellurium desorption proposed by Tinjod et al.\cite{Tinjod2003}. The thin CdTe layer was deposited by ALE (Atomic Layer Epitaxy), then the substrate was cooled down in the presence of tellurium flux in order to deposit amorphous tellurium, next the substrate was heated to growth temperature. Tellurium was evaporated and a ZnTe cap was grown.
Two series of samples were grown with two approaches for the deposition of a thin CdTe layer. In the first one we used a fixed number (12) of ALE cycles of CdTe which at 334 $^{\circ}$C corresponds to 3 monolayers (MLs) of CdTe and we varied the temperature during growth in order to influence QDs density. The low density of QDs was obtained when we strongly increased both deposition temperature of the CdTe layer and deposition temperature of ZnTe cap. In each case the deposition temperature of the CdTe thin layer and the ZnTe cap was the same. Second series of samples was obtained with various thickness of CdTe layer. This parameter varied from 4 to 16 ALE cycles of CdTe. During each of ALE loops 0.25 of monolayer has been deposited. Both the deposition temperature of CdTe layer and the ZnTe cap layer were fixed at 334${^\circ}$C. We found that samples with thinner CdTe layer exhibit lower QDs density in PL measurements.
For each of the described series of samples, the characteristic transformation of RHEED image from 2D to 3D, related to QDs formation\cite{Tinjod2003}, was observed (Fig.~\ref{fig:RHEEDimage}).

\section{Microphotoluminescence results}

\begin{figure}[!h]
\includegraphics[width=\linewidth]{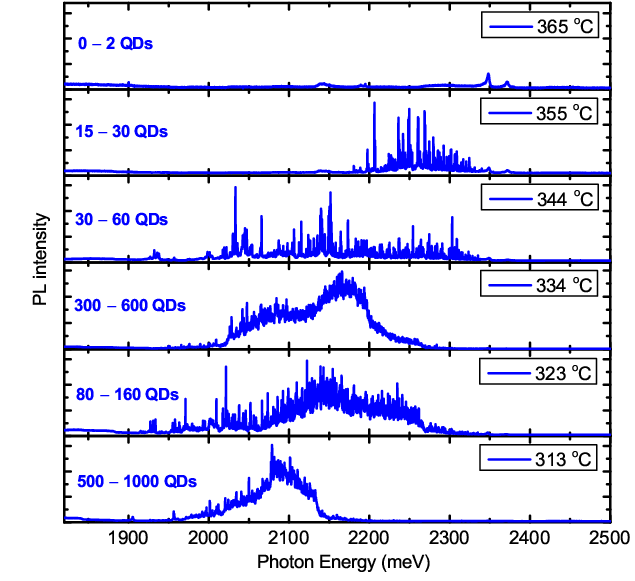}
\centering \caption[]{(color online) µPL spectra of CdTe/ZnTe QDs for various CdTe deposition temperature. Influence of the deposition temperature on the density of QDs and optical properties. Estimated numbers of QDs excited by laser spot (${\phi}=2$~${\mu}$m) are given.} \label{fig:uPLvariousT}
\end{figure}

\begin{figure}[!h]
\includegraphics[width=\linewidth]{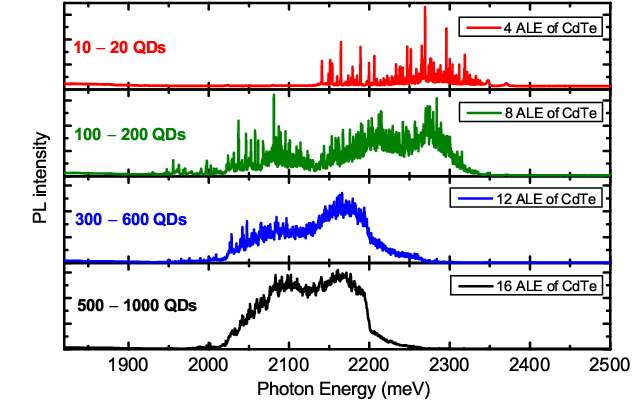}
\centering \caption[]{(color online) Influence of the CdTe layer thickness on the density of QDs and optical properties.    In this series of samples CdTe was deposited at 334 $^{\circ}$C. For this temperature 4 ALE correspond to 1 ML, 8 ALE to 2 MLs, 12 ALE to 3 MLs, and 16 ALE to 4 MLs of CdTe. Estimated numbers of QDs excited by laser spot (${\phi}=2$~${\mu}$m) are given.} \label{fig:uPLvariousALE}
\end{figure}

Low-temperature (T $=7$~K) microphotoluminescence (${\mu}$PL) measurements were conducted in typical setup with the microscope objective giving laser spot with about 2 ${\mu}$m diameter. Excitation was by a blue laser diode (405~nm). Spectral resolution was 0.08~meV. For both series of samples we observed sharp emission lines related to single QDs, which we consider as a confirmation of formation of QDs in all tested growth conditions (Figs. ~\ref{fig:uPLvariousT} and ~\ref{fig:uPLvariousALE}). The main difference between results obtained for various samples was the number of sharp PL lines observed in the spectrum. At the limit of the low excitation power each QD gives only a few intense lines (Figs. ~\ref{fig:3qds} and ~\ref{fig:PLanizotropy}): neutral exciton line (X) and one or two trion lines (X$^+$and/or X$^-$)\cite{Kazimierczuk2010}. This allows us to estimate the number of emitting dots per area of the laser spot, and consequently, the QDs density.
Example spectra are presented in Fig.~\ref{fig:uPLvariousT} for series of samples with different deposition temperature of CdTe layer. The increasing of CdTe deposition temperature results in decrease in the number of observed QDs lines. Also the averaged emission energy of the QDs ensemble is increasing. It shows that for higher growth temperature we observe smaller QDs density and the typical size of the dot is smaller or typical potential depth is more shallow. Such effects could be caused by temperature induced evaporation of CdTe layer. The decrease of the potential depth could be caused by temperature induced mixing of CdTe in QDs with ZnTe from the barrier. Depending on the temperature conditions we obtained samples with various estimated densities of QDs: with typical\cite{Tinjod2003} values of the density of QDs (10$^{10}$QD/cm$^2$), a sample with low-density of QDs (10$^9$QD/cm$^2$) and samples with ultra-low-density of QDs (10$^7$-10$^8$QD/cm$^2$).
We obtained very similar results for samples grown with density of QDs controlled by CdTe layer thickness (${\mu}$PL spectra presented in Fig.~\ref{fig:uPLvariousALE}). Decreasing the thickness of the CdTe layer results in reduction of density of QDs and increase of the averaged photon energy emission. We interpret this effect in the following way: thinner CdTe layer results in smaller size and number of QDs.

Both approaches to the reduction of QDs density: increasing CdTe deposition temperature and decreasing amount of deposited CdTe results in a similar evolution of optical properties of QDs ensemble (Figs.~\ref{fig:uPLvariousT} and~\ref{fig:uPLvariousALE}). We conclude that the key parameter in both cases is the amount of CdTe material overgrown by ZnTe cap. The main impact of the increased temperature during CdTe deposition is therefore desorption of CdTe material. We note here that both methods of controlling QDs density were combined in the method using amorphous tellurium\cite{Tinjod2003}. Other methods of CdTe/ZnTe QDs formation result generally in higher QDs densities\cite{Kobak2011} and reduction of the QDs requires separated study.

\section{New opportunities coming from low-density of dots}

\begin{figure}[!h]
\includegraphics[width=\linewidth]{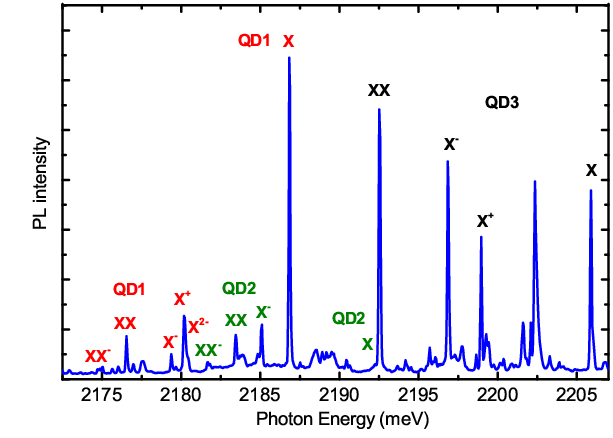}
\centering \caption[]{(color online) Low-density of QDs allows the identification of almost all emission lines in PL spectra. We identified emission lines from three QDs: neutral excitons (X, XX), and charged trions (X$^+$, X$^-$, X$^{2-}$,XX$^-$).} \label{fig:3qds}
\end{figure}

Reduction of the number of QDs allows easy identification (Fig.~\ref{fig:3qds}) and study of properties\cite{Kazimierczuk2011,Kazimierczuk2010} almost all emission lines in broad ${\mu}$PL spectrum. This is unavailable without using additional treatments like mesas in usual samples.

\begin{figure}[!h]
\includegraphics[width=\linewidth]{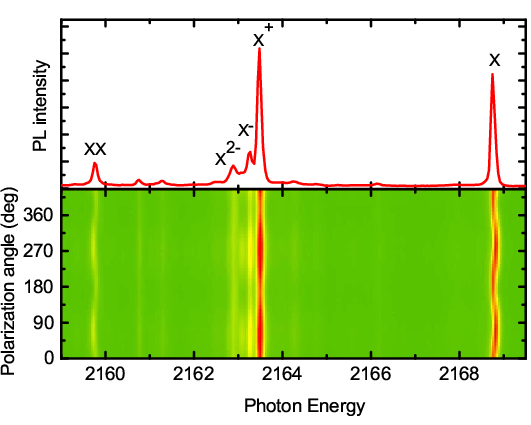}
\centering \caption[]{(color online) Emission lines show typical anisotropy properties: neutral exciton and biexciton have opposite linear polarization and trions are not linearly polarized.} \label{fig:PLanizotropy}
\end{figure}

The QDs grown by us with ultra low density exhibit typical properties of CdTe QDs: characteristic pattern of emission lines associated with different excitonic optical transitions in single QD. We identified strong lines related to neutral exciton in the highest energy, trions in intermediate energies, biexction in lowest energy, and weaker lines associated with higher charged states\cite{Kazimierczuk2011,Kobak2011,Kruse2011}. Emission lines were identified by a combination of various methods: measurement of luminescence intensity as a function of excitation power, linear polarization anisotropy measurements (Fig. ~\ref{fig:PLanizotropy}) and Zeeman effect.
Studying QDs with ultra low density opened for us the possibility to study QDs with various sizes and emission energies, including QDs with typical dimensions, with PL in the middle of the QDs ensemble.

\section{Conclusions}

We developed two methods of obtaining samples with ultra low density of QDs: increasing CdTe deposition temperature and decreasing CdTe layer thickness. Both methods give expected results - tunable QDs density. It is difficult to distinguish which method is better. Ultra low density of QDs allows the identification of almost all emission lines in whole PL spectra and the QDs show typical properties reported before for CdTe/ZnTe system. Successful control of QDs density opens new perspectives for spectroscopic studies of QDs.


\section{Acknowledgements}

We wish to acknowledge helpful discussions with Piotr Wojnar. This work was supported by Polish public funds in years 2011-2017 (NCN projects DEC-2011/01/B/ST3/02406 and DEC-2011/02/A/ST3/00131 and NCBiR project LIDER/30/13/L-2/10/NCBiR/2011). Project was carried out with the use of CePT, CeZaMat and NLTK infrastructures financed by the European Union - the European Regional Development Fund within the Operational Programme "Innovative economy" for 2007-2013.


\end{document}